\documentclass{article}

\usepackage{arxiv}
\usepackage{amsmath}
\usepackage[utf8]{inputenc} 
\usepackage[T1]{fontenc}    
\usepackage{url}            
\usepackage{booktabs}       
\usepackage{amsfonts}       
\usepackage{nicefrac}       
\usepackage{microtype} 
\usepackage{lipsum}
\usepackage{enumitem}
\RequirePackage[OT1]{fontenc}
\RequirePackage{amsthm,amsmath}
\usepackage{natbib}

\RequirePackage[colorlinks,citecolor=green,linkcolor=red, filecolor=magenta, urlcolor=blue]{hyperref}

\usepackage{graphicx,subfigure,amsmath,latexsym,amssymb}
\usepackage{upgreek,wrapfig}
\usepackage{comment}

\usepackage{algorithm,algpseudocode}
\usepackage{makecell}
\usepackage{listings}
\usepackage{url}
\usepackage{verbatim}

\newtheorem{theorem}{Theorem}

\newtheorem{as}{Assumption}
\newtheorem{lem}{Lemma}
\newtheorem{Proof}{Proof}
\newtheorem{rmk}{Remark}

\newcommand{\bg}{\mathbf{g}}
\newcommand{\Bz}{\mathbf{z}}
\newcommand{\Bw}{\mathbf{w}}

\newcommand{\bH}{\mathbf{H}}

\newcommand{\bN}{\mathbf{N}}

\newcommand{\bE}{\mathbf{E}}

\newcommand{\bz}{\mathbf{z}}

\newcommand{\bw}{\mathbf{w}}

\newcommand{\C}{\mbox{\boldmath{$\cal C$}}}

\newcommand{\e}{\ensuremath{\epsilon}}

\newcommand{\bm}{\mathbf}

\usepackage[usenames,dvipsnames]{xcolor}    
\newcommand{\myR}[1]{\textcolor{red}{#1}}
\newcommand{\myV}[1]{\textcolor{violet}{#1}}

\lstnewenvironment{rc}[1][]{\lstset{language=R}}{}

\title{Novel Bayesian   Procrustes Variance-based Inferences  in Geometric Morphometrics \\ \& Novel  R package: \texttt{BPviGM1}}

\author{

 Debashis ~Chatterjee\thanks{For suggestions and bug-report  regarding novel R package  \textbf{BPviGM1}, please send email to  \texttt{cdebashis.r@gmail.com}, or, please report new issue at \url{https://github.com/debashischatterjee111/BPviGM1/issues}} \\
  Interdisciplinary Statistical Research Unit\\
  Indian Statistical Institute\\
  Kolkata, India \\
  \texttt{debashis1chatterjee@gmail.com} \\

}

\begin{document}
\definecolor{bubbles}{rgb}{0.91, 1.0, 1.0}
\definecolor{blond}{rgb}{0.98, 0.94, 0.75}
\definecolor{beige}{rgb}{0.96, 0.96, 0.86}
	\definecolor{azure(web)(azuremist)}{rgb}{0.94, 1.0, 1.0}
	\definecolor{antiquewhite}{rgb}{0.98, 0.92, 0.84}
	\definecolor{floralwhite}{rgb}{1.0, 0.98, 0.94}
	\definecolor{ivory}{rgb}{1.0, 1.0, 0.94}
	\definecolor{lightcyan}{rgb}{0.88, 1.0, 1.0}
		\definecolor{snow}{rgb}{1.0, 0.98, 0.98}
		\definecolor{forestgreen(traditional)}{rgb}{0.0, 0.27, 0.13}
\maketitle

\begin{abstract}
	
  Classical Procrustes analysis   \citep{bookstein1997morphometric} has become  an indispensable tool under Geometric Morphometrics. Compared to abundant classical statistics-based literature, to date, very few Bayesian  literature exists on Procrustes shape analysis in Geometric Morphometrics, probably because of being a relatively new branch of statistical research and because of  inherent computational difficulty associated with Bayesian analysis.  On the other hand, although we can easily obtain the point estimators of the shape parameters using classical statistics-based methods, we cannot make inferences regarding the distribution of those shape parameters in general and, cannot put forward our prior belief and update to posterior belief on the same. We need to shift to Bayesian methods for that.   Moreover,  we may obtain a plethora of novel inferences from Bayesian Procrustes analysis of shape parameter distributions.  In this paper we propose to regard the posterior of Procrustes shape variance  as  morphological variability indicators, which is on par with various laws of mathematical population genetics like Hardy-Weinberg law \citep{stern1943hardy,masel2012rethinking}. Here we propose novel Bayesian methodologies for Procrustes shape analysis based on landmark data's isotropic variance assumption and propose a Bayesian statistical test for model validation of new species discovery using  morphological variation reflected in the posterior distribution of landmark-variance of objects studied under Geometric Morphometrics.  We will consider  Gaussian distribution-based and heavy-tailed t distribution-based models for Procrustes analysis. 

 To date, we are not aware of any direct R package for Bayesian Procrustes analysis for landmark-based Geometric Morphometrics. Hence, we introduce a novel, simple R package \textbf{BPviGM1} ("Bayesian   Procrustes Variance-based inferences in Geometric Morphometrics 1"),  which essentially contains the R code implementations of the computations for proposed models and methodologies, such as R function for Markov Chain Monte Carlo (MCMC) run for drawing samples from posterior of parameters of concern and R function for the proposed Bayesian test of model validation based on significance morphological variation.
 
 As an application, we applied our proposed Bayesian Procrustes Analysis on ``apes" data of  \cite{o1993sexual}  (also documented in R package ``shapes"). We compared the posterior variance of the shape of female vs. male for Gorilla, Chimpanzee,  Orangutan and conclude that there might be  ``Bayesian evidence"  in favor of a novel  hypothesis on face-shape: ``male primate manifests more fluctuation in face-shape than females," which suggests further research in future.
\end{abstract}

\small{\tableofcontents}
\newpage
\section{Introduction}
\label{sec:Introduction}
Procrustes shape analysis is one of the most important method for morphological identifications and morphological variation study. Morphometrics is the subject of the statistical study of biological shape and change of shape \citep{bookstein1997morphometric}.  Procrustes (Shape analysis) problems arise in a wide range of scientific disciplines, especially
when the geometrical shapes of objects are compared, contrasted \cite{theobald2006empirical}
and analyzed in \citet{theobald2006empirical}.   The shape is "the geometrical information that remains when location, scale, and rotational effects are filtered out from an object" \citep{kendall1977diffusion}. Two objects will have the same shape if one can be translated, rescaled, and rotated to the other to match exactly, in the sense that they are similar objects  \cite{micheas2006complex}. Geometric morphometrics is a discipline that focuses on the study of shape and morphological variations using statistical tools. In geometric morphometrics, the shape is usually diagnosed by collecting and analyzing length measurements, counts, ratios, and angles using Cartesian landmark and landmark coordinates capable of capturing morphologically distinct shape variables. Using classical Procrustes analysis  \cite{bookstein1997morphometric}, we can estimate shape parameters, but we will not get the posterior distribution of those. In principle, the Bayesian approach does not have such a problem because it provides multiple realizations to generate an "a-posteriori" distribution of parameters of the model \cite{fox2016applications}.
 
 Bayesian methods start with a prior belief about a parameter, thereby involving computation of update on belief after data is observed, known as ``posterior distribution". Posteriors involve a mathematical superposition of prior belief and evidence provided by observations. Hence,    Bayesian data analysis with suitable models offers a highly flexible, intuitive, and transparent alternative to classical statistics \cite{demvsar2020bayes4psy}.  We may obtain a wide range of novel inferences from Bayesian Procrustes analysis of shape parameter distributions, which we may not achieve if we stick to the classical statistics-based Procrustes approach.  For instance, in this paper, we propose to regard the posterior of Procrustes shape variance as morphological variability indicators. For example, inter-species and intra-species morphological variability may be reflected in Kulberk-Leibler divergence (KL-divergence) between the posterior and the respective variance parameters. KL-divergence is a divergence between distributions, and hence, the divergence between posteriors of shape-variance of different populations may be viewed as an indicator of morphological variability. We may note that we may not get such reasonable indicators merely from the euclidean distance between point estimates in general.  Moreover, such novel indicators involving posteriors of biometric shape-variance may reflect a natural extension of various laws of mathematical population genetics like Hardy-Weinberg law \citep{stern1943hardy,masel2012rethinking} (see section \ref{sec:KLdivPosterior}). Unfortunately, much of the modern era of science, Bayesian approaches remained on the sidelines of data analysis compared to classical statistical methods, mainly because computations required for Bayesian analysis are usually quite convoluted, often involving numerical calculation approximations of multi-dimensional integrals. Markov chain Monte Carlo (MCMC) methods are popular algorithms to sample efficiently from posterior, which we predominantly use in our paper.    
 
 In this paper, we propose novel Bayesian methodologies for Procrustes shape analysis based on landmark data's isotropic variance assumption and propose a Bayesian statistical test for model validation of new species discovery using inter-species intra-species morphological variation reflected in the distribution of landmark-variance of objects studied under Geometric Morphometrics. We will assume isotropy of variance for landmark data, i.e., the nonsingularity of variance-covariance matrices.
 
 For an application to real geometric morphometric data, we applied our proposed Bayesian Procrustes Analysis methodology on ``apes" data of  \cite{o1993sexual}  (documented in R package ``shapes"). In particular, under isotropic variance assumption (Assumption \ref{assumption:Isotropy}), We compared the posterior of variance of shape of female vs. male for different primates (Gorilla, Chimpanzee, Orang utang) and come to a conclusion that there might be  ``Bayesian evidence"  in favor of a novel biometric hypothesis on shape: ``male primate manifests more fluctuation in face-shape than females", which suggests further research in future.
 
At present, there is no R package for Bayesian Procrustes shape analysis using landmark-based objects as data, at least not that we know to date. Hence,  we  combine the whole computational parts of this paper into a novel, simple and flexible R package \textbf{BPviGM1}("Bayesian Procruste Analysis using  Landmark based on Isometric or Anisometric Error under Geometric Morphometrics"), which will implement the novel bayesian models and methodologies introduced in this paper. 


The rest of the paper is organized as follows.  Section \ref{sec:Introduction} contains the Introduction part as well as overview,section \ref{sec:KLdivPosterior} contains existing  literature review, section \ref{sec:preliminary} contains preliminaries on Landmark based Object Analysis, Kendal’s Shape Space, Principle of Procrustes Shape Analysis  and overview of principle behind  Classical Method for Procruste Analysis. Section \ref{sec:NovelBayesian} contains our proposed novel Bayesian Models \& methods for Procrustes Analysis under Isotropic Error Variance, along with sampling Procedure ( Markov Chain Monte Carlo Approach for posterior sampling), which we illustrate using Simplest 2D Regression Modelwith Isotropic Landmark Variance and both uniform prior and  Empirical prior with a detailed discussion on problem of Ill-Posedness \& Remedy using Empirical Prior. The theory behind our proposed novel measure of intra-inter-species shape variability in subsection \ref{subsec:ill-posedness}. Section \ref{sec:Application} contains  Applications of our proposed methodologies on 
\begin{enumerate}
     \item Convergence diagnostics using simulated random convex-concave polygon in 2D,
    \item Bayesian Procrustes Analysis on “apes” Data: fluctuation in face-shape for Male vs. Female,
    \item  Novel Bayesian Procrustes Analysis on Paleontological Objects : Trilobite shape data (2D Landmark) \& Dinosaur bone-shape data (3D Landmark),
    \item Application of Novel Measure of intra-species and Inter-species Shape Variability on “apes” data,
    \item Over-fit Problem using Objects already in Shape-Space.
\end{enumerate}
 
section \ref{sec:RpackageSourcecode} contains details of installation procedure for novel R Package \texttt{BPviGM1} \& Source Code availability for the same and, main functions described in our R package. We conclude in section \ref{sec:Conclusion}.

\section{Literature Review}
\label{sec:Literature Review}
The book of  \cite{bookstein1997morphometric} contains  systematic survey of morphometric methods and shape analysis for landmark data using conventional multivariate statistical analysis, solid geometry and  biomathematics for  biological insights into the features of many different organs and organisms. \cite{klingenberg2002shape} contains methodologies for analysis of symmetric structures using  quantification of  variation among  objects. \cite{rohlf1990extensions} contains a review and some generalizations of superimposition methods for comparing configurations of landmarks in two or more specimens. Use of  Procruste-Variance is common in literature for  study of morphological variation, e.g.,\cite{stange2018study}.  Landmark-based geometric morphometric study has becoming increasingly popular and the popularity of statistical methodology research regarding shape analysis  is tending to shift from classical to bayesian.  
 Although there are plenty of classical statistics based literature  which addresses various aspects of shape analysis based on landmark and semi-landmark under geometric morphometrics, bayesian  research on the same is comparatively  new and till date, there are very few bayesian literature on the same. Nevertheless, bayesian methods  becoming increasingly popular and there are some recent advancements on the  bayesian  theory of shape analysis like    \cite{theobald2006empirical}, \cite{fox2016applications}, \cite{gutierrez2019bayesian}, \cite{micheas2010bayesian}, although most of the existing  bayesian literatures depend heavily on isotropic error variance of the landmark data. In this paper we will generalize the existing methods and propose  novel models and methods to handle even the anisotropic error variance using Mahalanobis distance of \cite{mahalanobis1925analysis}, \cite{mahalanobis1936generalized}. Interestingly, Mahalanobis's definition was prompted by shape analysis and shape comparison problems \citep{mahalanobis1925analysis}.


\section{Preliminaries}
\label{sec:preliminary}

\subsection{Landmark based Object Analysis}
Landmarks are point locations that are biologically and morphologically homologous between specimens \citep{gunz2013semilandmarks}. We will consider a particular object with a finite number $k$ of points in $d$ dimensions (For instance, $d=2$ or $d=3$ for two or three dimensions, respectively).  We often select these points on complex objects' continuous surface, such as a fossilized dinosaur bone. 
Sometimes too few landmarks are available, or a situation occurs when some structures cannot be quantified using traditional landmarks. For example, a traditional homologous landmark can not capture the shape of visible muscle attachments on bones. In such a scenario, semi landmarks make it possible to quantify two- or three-dimensional homologous curves and surfaces and analyze them together with traditional landmarks \citep{gunz2013semilandmarks}.  We can digitize the outlines as a series of discrete points with the individual points that must be slid along a tangential direction to remove tangential variation \citep{perez2006differences}, known as semilandmarks or sliding landmarks. 

A 2D landmark or 3D landmark data can be represented by 3D array $(p \times k \times n)$, which is the
required input format for many functions in package geomorph (see section). where
\begin{center}
\begin{description}[noitemsep] 
	\item[$p=$]  the number of landmark points,
	\item [$d=$] the number
	of landmark dimensions (2 or 3 for two or three dimensions, respectively),
	\item [$n=$] the number of specimens.
\end{description}
\end{center}
For Procrusts analysis, we cannot directly work with raw  landmark data. First, we need to transform into Kendal's Shape space \citep{dryden1998statistical, dryden2016statistical} (see Subsection \ref{subsec:KendalShapeSpace}).

For a 2D  landmark, we can equivalently represent a particular landmark with coordinate $(x, y)$ with a complex number $z=x+iy$, where $x$ is the real part, and $y$ is the imaginary part. 
The plane of matrices $M(2, \Re)$ generated by
${\displaystyle {\begin{pmatrix}x&y\\-y&x\end{pmatrix}}}$, 
is isomorphic to the complex number plane $\C$, and the  rotation matrix $R_{\theta}:={\displaystyle\begin{pmatrix}
	\cos\theta & -\sin\theta \\
	\sin\theta & \cos\theta
	\end{pmatrix}}$ in 2D is equivalent to multiplication by $e^{i\theta}$  in complex plane $\C$, which acts on the plane as a rotation of $\theta$ radians. Hence, using complex number notation,  the rotation operation by angle $\theta$  can be  represented easily using  multiplication by a factor $\exp(i\theta)$, which is equivalent to a rotation matrix. For 3D landmark, we cannot directly use complex number and have to make use of 3D rotation matrix (see Appendix \ref{Appendix}).



\subsection{Kendal's Shape Space}
\label{subsec:KendalShapeSpace}
Shape is the geometry of an object modulo position,
orientation, and size. Mathematically, A shape is a point in a high-dimensional, nonlinear
manifold, called a shape space. For preliminaries about manifold, see Appendix \ref{Appendix}. We require a  metric space structure  for a comparison
between two shapes. Each shape determined by a set of landmarks can be represented by a point in Kendall’s shape space.

\begin{figure}[H]
\centering
\begin{minipage}{.48\textwidth}
  \centering
  \includegraphics[width=\linewidth]{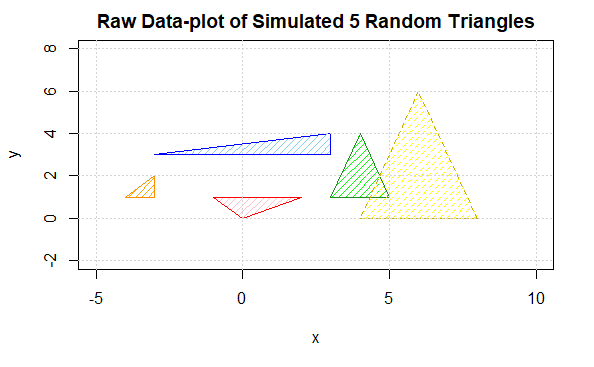}
  \caption{Plot of 5 triangles of randomized coordinates. This is an illustration of objects (triangles with vertices as landmarks) in configuration space $\Re^{3\times2}$. Here the landmark coordinates (vertices) are taken from 2D uniform distribution, except the orientation of red triangle has made different, which shows red triangle forms different shape from the rest (reflection is not included)}
  \label{fig:rawtriangle}
\end{minipage}%
\hspace{0.1cm}
\begin{minipage}{.48\textwidth}
  \centering
  \includegraphics[width=\linewidth]{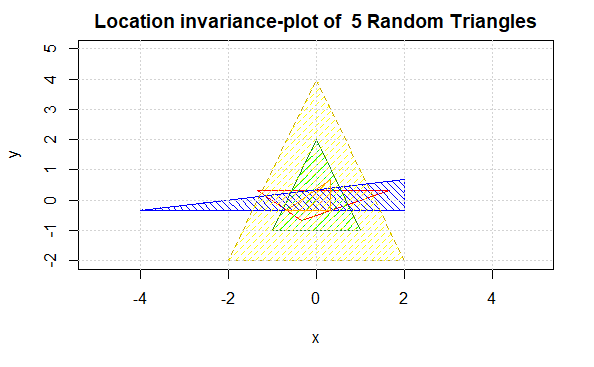}
  \caption{Plot of location shifted (not scaled)   space (after translation, scale  invariant applied  using Helmert matrix in comparison to raw data plot of triangles (figure \ref{fig:rawtriangle}). Observe that,  green \& yellow triangle pair are now almost overlapped (if scaled)  because of similarity of their shapes.}
  \label{fig:preshapesp}
\end{minipage}
\end{figure}

\begin{figure}[H]
  \centering
  \includegraphics[width=0.9\linewidth]{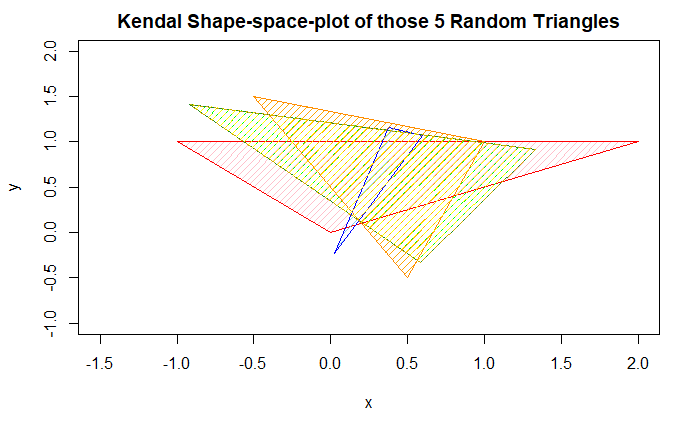}
  \caption{Plot of  shape space (using classical Procrustes Analysis in comparison to raw data plot of triangles (figure \ref{fig:rawtriangle}) and also in comparison to pre-shape space plot (figure \ref{fig:preshapesp}). Observe that, blue \& orange triangle-pair are NOT merged  (and also green \& yellow triangle pair NOT merged), which is due to the fact that Landmarks were deliberately inputted counter-clock wise, which accounts for the upside-down flip of blue triangle in comparison to orange triangle.}
  \label{fig:CFPF_shapesp}
\end{figure}

\begin{figure}[H]
\centering
  \centering
  \includegraphics[width=0.9\linewidth]{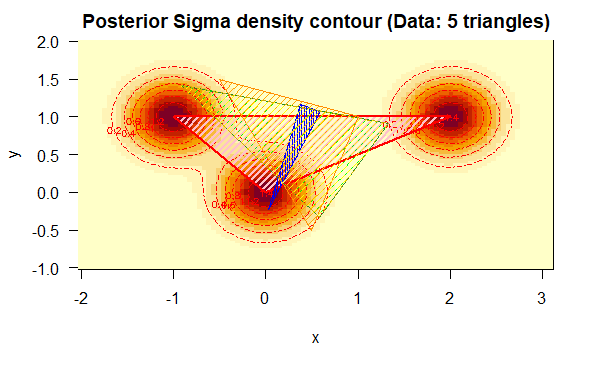}
  \caption{Plot of Posterior density contour using 5 triangles of randomized coordinates, using methods discussed in section \ref{sec:NovelBayesian}. Observe that, the blue triangle is of different shape which has been depicted from the plot as it predominantly  remains outside of inner dense-coloured contour. }
  \label{fig:Bayes_CFPF_shapesp}
\end{figure}

The data for raw objects can be arranged such that
the coordinates of each landmark are found on a separate row, or that each row contains all landmark
coordinates for a single specimen. Following \cite{dryden1998statistical}, we can mathematically express  at least three spaces on which models can be defined 
\begin{description}
	\item[Configuration space:] The space for raw objects  representation. Formally, The Euclidean space $\Re^{p\times d}$, where
	configurations $\bz_1,\cdots, \bz_n \in \Re^{p\times d}$ are  d-dimensional shapes such that $\bz_i=\left(z_{i1}, z_{i2}, \cdots, z_{ip}\right)^T$  are vector containing  $p$ many  landmarks coming from $\Re^d$, for $i =1, 2, \cdots, n$, where $T$ denotes transposition operator. For 2 dimension, $d=2$ and can equivalently represented by complex plane.
		\item[Pre-shape space:] The space obtained from configuration space after  translations and scalings have been removed. Formally, it is  represented by a hypersphere of unit radius in $d(p-1)$ dimensions. We can go from the configuration space to the pre-shape space using
   the transformation    
   \begin{eqnarray}
   \bz^{+} = \dfrac{H\bz}{\|H\bz\|},
   \end{eqnarray}
   where $H$ is the Helmert sub-matrix   \citep{lancaster1965helmert,dryden1998statistical}. In particular, examples of Helmert matrix includes many orthogonal matrices and   rotation matrices in 2D and 3D, many of which will be used in this paper (see subsection \ref{subsec:Helmert} in Appendix \ref{Appendix}). 
  \item[Shape space:] Space for which configurations are invariant under location, rotation, and scaling. We can go from the configuration space to the shape space via Procrustes Analysis. The method might be classical or Bayesian.
\end{description}
Figure \ref{fig:rawtriangle} shows triangles (objects) in configuration space, whereas figure \ref{fig:preshapesp} shows those objects in pre-shape space (after making translation \& scale invariance). Figure \ref{fig:CFPF_shapesp} shows the corresponding shape space (using classical full Procrustes analysis). Lastly, figure \ref{fig:Bayes_CFPF_shapesp} shows the Bayesian version (using a novel methodology of this paper, see section \ref{sec:NovelBayesian}), which shows the posterior contour of Landmark variance. Intuitively, the blue triangle shape is different from others because one of its side is very small relative to its other two sides, whereas all other triangles have roughly proportional sides. As we can infer from the posterior variance-plot using methodologies discussed in section \ref{sec:NovelBayesian}, blue triangle fails to capture  the contour's sufficient weight, indicating a probable outlier-object.

\begin{figure}[H]
\centering
\begin{minipage}{.48\textwidth}
  \centering
  \includegraphics[width=\linewidth]{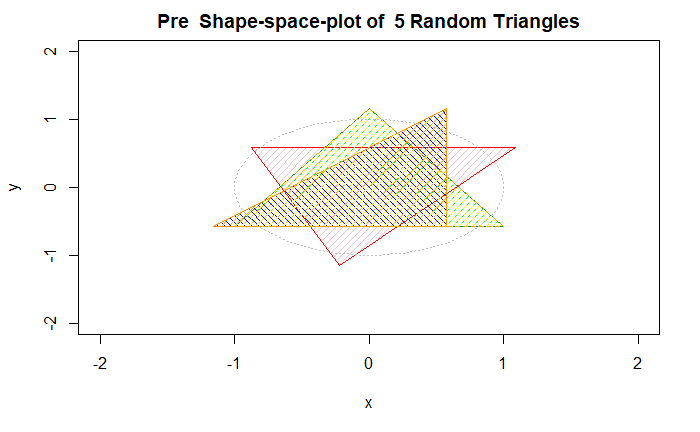}
  \caption{Plot of 5 triangles of randomized coordinates in Pre-shape space(translation \& scale invariance).  Here the landmark coordinates (vertices) are taken from 2D uniform distribution, except the orientation of red triangle has made different, which shows red triangle forms different shape from the rest (reflection is not included)}
  \label{fig:preshapedtriangle}
\end{minipage}%
\hspace{0.1cm}
\begin{minipage}{.48\textwidth}
  \centering
  \includegraphics[width=\linewidth]{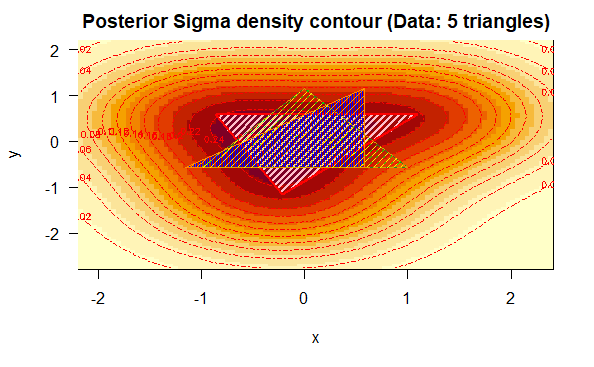}
  \caption{Plot of posterior-contour of Landmark variance parameter $\sigma$ using novel Bayesian Procrustes Analysis of section \ref{sec:NovelBayesian}. Here the rotation has not been considered. Those 5 triangles of randomized coordinates in Pre-shape space(translation \& scale invariance). }
  \label{fig:preshapesp}
\end{minipage}
\end{figure}

\section{Principle of  Procrustes Shape Analysis}
The word ``Procrustes''refers to a bandit from Greek mythology who made his victims fit his bed either by stretching their limbs or cutting them off.   In its most
general formulation, Procrustes analysis involves the optimal
matching of two or more form matrices. Here we consider only sets
of forms where each form matrix has the same dimensions. 
 The definition of
“shape” \citep{kendall1977diffusion} is “all the geometrical information that remains when location, scale and rotational
effects are filtered out from an object”. 
The word “scale” can be regarded as positive scalar $s$ satisfying the equation $g(s \cdot z) = s\cdot g(z)$, where $g(X)$ represents any positive
real-valued function of the complex  vector $z$ or (equivalently configuration matrix).

  2-dimensional shapes is defined by a set of $p$ landmarks represented by a
complex vector $ \Bz = (z_1, z_2, \cdots, z_p)^T$ 
where the real part of each $z_i$ is the x-coordinate and the
imaginary part is the y-coordinate.  \cite{fox2016applications}  considered matching cells based on either their location (centroid
based), or their shape differences (Procrustes matching).

For the Full Procrustes Fit (FPF) an object is translated, rotated and dilated to produce an exact
match with another object. 

Consider two  configurations  $\Bw = (w_1, w_2,\cdots, w_p)^T$  and  $\Bz = (z_1, z_2,\cdots, z_p)^T$ both in $\Re^{d\times p}$ (alternatively, using complex plane for 2D landmark, both in $\C^p$), where $\bz$ is usually  centered, i.e., $\Bz^\star \mathbf{1}_p =0$, ($\Bz^\star$ denotes the transpose of the complex conjugate of $\Bz$and $\mathbf{1}_p$ is the $p\times 1$ vector of ones). In order to compare the configurations in shape we need to establish a measure of distance between the two shapes \citep{dryden1998statistical}. 
 The object is to match  to a mean object $\Bz$  using the complex regression (or equivalently, multivariate regression for 2D or 3D euclidean coordinate). For   scalar dilation parameter $b$, $c \in \Re^d$ (for example, $c=(c_1, c_2)$ for 2D landmark or equivalently, for complex plane $c=c_1+ic_2$), and for $\bw, \bz \in \Re^{d \times p}$ we consider the multivariate  regression equation 

\begin{eqnarray}
\label{multiple_regression_1}
\Bw_{d\times p} &=c_{d\times 1}^{T}\mathbf{1}_p+ bR_{\theta}\Bz_{d\times p} +\epsilon_{d\times p}
=\left[\mathbf{1}_p, \Bz\right]A+\epsilon
=X_D A+\epsilon,
\end{eqnarray}
or, equivalently,

\begin{eqnarray}
\label{multiple_regression_1M}
\Bw &=\left[\mathbf{1}_p, \Bz\right]A+\epsilon
=X_D A+\epsilon,
\end{eqnarray}
where the matrix containing parameters is  is the $(d+1)\times d$ dimensional $A=\left[c, bR_{\theta}\right]^T$ (using rotation matrix notation $R_{\theta}$),    $X_D:=\left[\mathbf{1}_p, \Bz\right]$ is the $ (d+1)\times p$ 'design matrix', $c$ is the $d$ dimensional vector corresponding to  translation, $b$ is the complex number corresponding to  the dilation, and $\theta$ is the rotation angle and $\epsilon$ is a $d \times p$ -dimensional complex
error vector. For our convenience, we will omit the notations for dimension.
Under isotropic error variance assumption \citep{fox2016applications}, where it is assumed that the variations of the landmarks are independent, The popular distribution for $\epsilon_{d\times p}=(\epsilon_1, \epsilon_2, \cdots, \epsilon_d)^T$ (each $\epsilon_i \in \Re^d$) will be 
\begin{eqnarray}
\epsilon_i \sim \bN_{p}(0, \sigma^2\mathbb{I}_{d}),
\end{eqnarray}
where $ \bN(\cdot)$ is multivariate normal distribution \citep{goodman1963statistical}(see Appendix) with $\mathbb{I}_{p}$  being the identity matrix. In practice, assumption of isotropic error variance may not hold and in this paper we will not assume isotropy. In this paper we will assume 

\begin{eqnarray}\label{Our_paper_error}
\epsilon \sim \bN_{d\times p}(0, \Sigma_{dp\times dp})
\end{eqnarray}
where $\Sigma_{dp\times dp}$ is the variance-covariance matrix of $\epsilon$, which in general, represent the dependency structure among the landmarks (which  almost always happens in reality).

\paragraph{Special case (using complex plane):}
If we use complex plane, then we can rewrite \eqref{multiple_regression_1} as 
\begin{eqnarray}
\label{complex_regression_1}
\Bw&=c\mathbf{1}_p+ be^{i\theta}\Bz +\epsilon
=\left[\mathbf{1}_p, \Bz\right]A+\epsilon
=X_D A+\epsilon,
\end{eqnarray}
where $A=\left[c, be^{i\theta}\right]$ (equivalently, $A=\left[c, bR_{\theta}\right]$ is the $d\times (d+1)$ dimensional location-dilation-rotation matrix using rotation matrix notation $R_{\theta}$ in place of $e^{i\theta}$) is the matrix containing parameters and  $X_D:=\left[\mathbf{1}_p, \Bz\right]$ is the $ (d+1)\times p$ 'design matrix', $c$ is the complex number corresponding to  translation, $b$ is the complex number corresponding to  the dilation, and $\theta$ is the rotation angle and $\epsilon$ is a p-dimensional complex
error vector. Under isotropic error variance assumption \citep{fox2016applications}, where it is assumed that the variations of the landmarks are independent, The popular distribution for $\epsilon$ is
\begin{eqnarray}\label{err_full}
\epsilon \sim \C\bN_{p}(0, \sigma^2\mathbb{I}_{p\times d}),
\end{eqnarray}
		where $\C \mathcal N(\cdot)$ is complex normal distribution \citep{goodman1963statistical}(see Appendix) with $\mathbb{I}_{p}$  being the identity matrix. In practice, assumption of isotropic error variance may not hold and in this paper we will not assume isotropy. In this paper we will assume 
		
		\begin{eqnarray}\label{Our_paper_error}
		\epsilon \sim \C\mathcal N_{p}(0, \Sigma)
		\end{eqnarray}
		where $\Sigma$ is the variance-covariance matrix of $\epsilon$. For $d=2$ and complex plane, $\Sigma =\Sigma_1+i\Sigma_2$. 
	


\begin{as}[Isotropy of variance parameter]
\label{assumption:Isotropy}
    For   the general Bayesian regression model  \ref{multiple_regression_1M},   variance for landmark-based objects is isotropic, i.e., the   variance-covariance matrices $\Sigma$ of \eqref{err_full} is nonsingular. In particular, the landmark variances are independent.
\end{as}

Assumption \ref{assumption:Isotropy} is needed to by pass the pathological scenario of having more unknown parameters than number of observations, known as ``ill-posed" problem (refer to subsection \ref{subsec:ill-posedness}).

\subsection{Classical Method for Procruste Analysis }
\label{subsec:Calssicalmethod}
Existing bayesian literature like \cite{fox2016applications} has assumed that, in equation \eqref{complex_regression_1}, $\epsilon$ is a p-dimensional complex error vector  to have an isotropic covariance matrix, i.e., any direction for a landmark is
equally preferred. In other words, for isotropic case, we can model 
To carry out the superimposition by classical statistical method we  estimate $A$ by minimizing the least squares objective function, the sum of square of errors 

\begin{eqnarray}
\label{classical_SSE}
D^{2}_{iso}\left(\Bw, \Bz\right)= \epsilon^\star \epsilon&=\left(\Bw-X_D A\right)^\star \left(\Bw-X_D A\right)=\| \bw-c\mathbf{1}_p- be^{i\theta}\bz  \|^2,
\end{eqnarray}

The classical statistical idea is to find the FPF of $\Bw$ onto $\Bz$ that minimizes the magnitude of the difference in shape
between $w$ and $z$.  Following \cite{dryden1998statistical}, we can rewrite in terms of FPD (full Procrustes distance between complex configurations $\Bw$ and $\Bz$)
 is



\begin{eqnarray}
\label{Full Procrustes Distance_iso}
D^{2}_{F, iso}\left(\Bw, \Bz\right)= \inf_{b,\theta,c}D^{2}_{iso}\left(\Bw, \Bz\right),
\end{eqnarray}
which shows the magnitude of the difference in shape
between the two objects \citep{dryden1998statistical}. The classical full Procrustes superimposition of $\Bz$; on $\Bw$ is obtained by estimating $A$ with $\hat{A}$, where 
  
  \begin{eqnarray} \label{classical_criteria}
 \hat{A}=\arg\min \epsilon^\star \epsilon.
  \end{eqnarray}

\begin{rmk}[usefulness of Bayesian method over Classical]
	\label{usefulness_of_bayesian}
	Classical (frequentist) statistical method-based Procrustes analysis  has its own advantages and disadvantages.
	Although it is easy to apply and obtain the point estimators of the shape parameters, their sampling
	distribution cannot be obtained in general \citep{fox2016applications}, \citep{micheas2010bayesian}. In other words, in general from classical approach it is not possible to make additional inference on the parameters except  the point estimates. We cannot obtain in general the
	distributions of these estimates, even under the case of easiest possible error distribution (normally  distributed error models) \citep{micheas2010bayesian}. Moreover,  from classical approach 
	 the distribution of the full Procrustes distance is also, in general hard to work with, even under popular multinormal models \citep{dryden1998statistical,  fox2016applications,micheas2010bayesian}.
\end{rmk}


\section{Novel Bayesian Models \& methods  for Procrustes Analysis}
\label{sec:NovelBayesian}

The principle of Bayesian Procrustes Shape Analysis under isotropic error variance for landmark data  adapted in this paper is a novel generalization  of principle stated in \cite{fox2016applications}, \cite{micheas2010bayesian}, whereas novel models and methodologies has been proposed under anisotropic error variance using Mahalanobis distance.

Bayesian regression fit estimate does not always match the least-square estimate of the classical method \eqref{Full Procrustes Distance_iso}, rather it minimizes Bayes risk. let $L(\theta,\widehat{\theta})$ be a loss function, such as squared error (see appendix). The  expected loss of  an estimator  $\widehat{\theta}$ of parameter $\theta$ is defined as $E_\pi(L(\theta, \widehat{\theta}))$, where the expectation is taken over the probability distribution of  $\theta$. Now, $\widehat{\theta}$ will be  a ''Bayes estimator'' if it minimizes the Bayes risk (the posterior expected loss) among all estimators.

Hence, the aim of Bayes estimtion is to choose the most appropriate  estimator $\widetilde{\lambda}=(\widetilde{b}, \widetilde{c}, \widetilde{\theta}, \widetilde{\sigma})$ among all estimators  $\widehat{\lambda}=(\widehat{b}, \widehat{c}, \widehat{\theta}, \widehat{\sigma})$ of the unknown parameter $\lambda=(b,c,\theta, \sigma)$ such that

\begin{eqnarray}
  \widetilde{\lambda}= \displaystyle \arg\min_{\widehat{\lambda}} E(L(\lambda,\widehat{\lambda}) | \bw, \bz).
\end{eqnarray} 
The most common risk function used for Bayesian estimation is the mean square error (MSE). This is   a Bayesian version of classical criteria \eqref{classical_criteria}, known as ''squared error risk''. The $MSE_B$ is defined by
\begin{eqnarray}
\label{Bayesian_criteria}
\mathrm{MSE}_B = E\left[ (\widehat{\lambda} - \lambda)^2 \right],
\end{eqnarray}

where the expectation is taken over the joint distribution of $(b,c,\theta, \sigma)$ and $\bw$.
We will use  the MSE of \eqref{Bayesian_criteria} as risk, which implies that the Bayes estimate of the unknown parameter is simply the mean of the posterior distribution.

\subsection{Bayesian Model for  Procrustes analysis  under  Isotropic Error Variance}
Bayesian Procrustes analysis  starts with idea  that  the observed values of $\Bw$
can be represented as  samples from a model distribution with a property that the  highest probability of realization of a sample  is around the mean value $\bz$.

\begin{as}[Essential Property of  Error Density  for Bayesian Model  (under isotropic  variance)]
	\label{Ass_Intuition-1}
	Selection of Bayesian model  for Bayesian  Procrustes analysis  with density of each random variable ( unobserved random variable $\overrightarrow{\bw}$ corresponding to the observed value $\bw$) should be such that,  it has its peak value (global optimum) at unobserved truth value of parameters   $ c_0\mathbf{1}_p+ b_0e^{i\theta_0}\bz$ and the error is such that the probability of a typical sample  to fall inside a $r-$ ball around $\bz$ is higher than  to fall inside a similar  $r-$ ball around  some other $\tilde{\bz} \neq \bz$. In other words, bayesian density criteria (complementary to classical criteria \eqref{classical_criteria}) can be stated as 
	
	\begin{eqnarray}
	\label{bayes density criteria}
	P\left(\left\lbrace \bw : \| \bw-c_0\mathbf{1}_p- b_0e^{i\theta_0}\bz  \| \leq r\right\rbrace \right) \geq P\left(\left\lbrace \bw : \| \bw-c\mathbf{1}_p- be^{i\theta}\tilde{\bz}  \| \leq r \right\rbrace \right)  , \ \ \ \forall \ \ r >0, \  \  \forall \  (c,b,\theta,\tilde{\bz}) \neq (c_0, b_0,\theta_0,\bz).
	\end{eqnarray}

\end{as}

To date, most of the existing bayesian literature on shape analysis like \citep{fox2016applications, micheas2006complex}  have assumed  isotropic error variance for landmarks and assumed bayesian models which satisfies assumption \ref{Ass_Intuition-1}. Here we will also 
 propose two  Bayesian models which satisfies assumption \ref{Ass_Intuition-1}, under isotropic error variance.

\begin{eqnarray}
\label{bayesian_density_A}
\text{Model A:} \ \    f_{\bz}(\Bw|\cdot) \propto \sigma^{-1} \exp\left(\sigma^{-2}D^{2}_{iso}(\bw,\bz)\right).
\end{eqnarray}

\begin{theorem}\label{theorem_1}
	Density stated in Model \eqref{bayesian_density_A} satisfies Assumption \ref{Ass_Intuition-1}  for large numbers of landmarks $p$ and large numbers of objects $n$, if the truth model is  distributed with mean $c_0+b_0e^{i\theta_0}$ and variance $\sigma_{0}^{2}\textbf{I}_{pn}$. In other words, for suitable prior density $\pi(c,b,\theta,\sigma)$ if posterior consistency holds in the sense 
		\begin{eqnarray}
	\label{eq:theorem_if1_posteriorConv}
\displaystyle\lim_{r\rightarrow 0}\left(\lim_{p,n\rightarrow\infty}\displaystyle \int_{\left\lbrace \bw : \| \bw-c\mathbf{1}_p- be^{i\theta}\tilde{\bz}  \| \leq r\right\rbrace} f_{\tilde{\bz}}(\Bw|c,b,\theta,\sigma) \quad d\pi(c,b,\theta,\sigma)\right) =  f_{\bz}(\Bw|c_0,b_0,\theta_0,\sigma_0),
	\end{eqnarray}
	
then,	for all $\tilde{\bz}\neq \bz$ and for all $r>0$
		\begin{eqnarray}
	\label{eq:theorem_1}
\displaystyle\lim_{p,n\rightarrow\infty} \int_{\left\lbrace \bw : \| \bw-c_0\mathbf{1}_p- b_0e^{i\theta_0}\bz  \| \leq r\right\rbrace} f_{\bz}(\Bw|c,b,\theta,\sigma) \quad d\pi(c,b,\theta,\sigma)		\geq\displaystyle \int_{\left\lbrace \bw : \| \bw-c\mathbf{1}_p- be^{i\theta}\tilde{\bz}  \| \leq r\right\rbrace} f_{\tilde{\bz}}(\Bw|c,b,\theta,\sigma) \quad d\pi(c,b,\theta,\sigma),
	\end{eqnarray}
	where $f_{\tilde{\bz}}(\Bw|c,b,\theta,\sigma)$ is as stated in \ref{bayesian_density_A}.
\end{theorem}

Based on assumption \ref{Ass_Intuition-1} we can generalize the density \eqref{bayesian_density_A} with

\begin{eqnarray}
\label{bayesian_density_general}
\text{ Model AG:} \ \ f(\Bw|\cdot) \propto  \bg\left(\sigma, D^{2}_{iso}(\bw,\bz)\right),
\end{eqnarray}
 such that the function $\bg(\cdot)$  has global maxima at $\bz$.

 \subsection{Sampling Procedure:  Markov Chain Monte Carlo Approach}
 
 Sampling for posterior is one of the main challenges for Bayesian Procrustes analysis  because often the full form of prior or even the likelihood will not be known. There are popular sampling methods  to handle these computational difficulties. In this paper, we will resort to {\em MCMC (Markon Chain Monte Carlo)} based method because of its flexibility general applicability. The {\em Gibbs sampler} is a special case of the Metropolis-Hastings algorithm and adopted in existing bayesian literature like \cite{fox2016applications}, but it differ in two ways: first, we always accept candidate point  and secondly, we  Need to know full conditional distributions.
 For both models \eqref{bayesian_density_A}, \eqref{bayesian_density_general} respectively, we can sample the parameters using Gibbs sampler only when full conditional distributions is known, for example, when the landmarks are in pre-shape space and only parameters to be inferred are the variance-parameters with non-informative priors (refer to Algorithm \ref{Algorithm_Gibbs} in Appendix \ref{Appendix}.  In general,  is often not possible to get full expressions of all conditional distributions.  In particular, for empirical prior or for anisotropic  landmark variance Gibbs sampler may not work atall  and MCMC (Markon chain Monte Carlo, a version of Metropolis Sampling) is necessary. The outline of simple version of MCMC is as follows: after choosing initial values of $(b, c, \theta, \sigma)$  arbitrarily, we perturb the parameter point and calculate the density ratio, we accept a new parameter point if the density ratio becomes greater than unity, in fact we accept a new sample parameter point with density-ratio-dependent acceptance probability. the detailed algorithm is stated in algorithm \ref{Algorithm:MCMC_iso}.  
 
   We will allow for a burn in period of $1000$, we will run the sampler for $20000$ samples and then take the sample mean, which will give us the minimizer of the Bayes MSE \eqref{Bayesian_criteria} (Bayes estimator), which we denote as $\widetilde{\lambda}=(\widetilde{b}, \widetilde{c}, \widetilde{\theta}, \widetilde{\sigma})$.
 Then the BFPF (Bayesian Full Procruste Fit) will be 
 
 \begin{eqnarray}
 \label{BayesianFPF_multiple}
 \Bw_{BFPF} &=\widetilde{c}\mathbf{1}_p+ \widetilde{b}R_{\widetilde{\theta}}\Bz.
 \end{eqnarray}
 Alternatively, if we write using complex number (for 2D landmark), then
 \begin{eqnarray}
 \label{BayesianFPF_Complex}
 \Bw_{BFPF}&=\widetilde{c}\mathbf{1}_p+ \widetilde{b}e^{i\widetilde{\theta}}\Bz
 \end{eqnarray}
 
 \subsection{Illustration: Simplest 2D Regression Model: with Isotropic Landmark Variance and Empirical prior}
 For illustration, as a special case of \eqref{complex_regression_1}, with number of objects being $n$.  
 Here we will assume that we are in pre-shape space, i.e., both $\bw$ and $\bz$ has been transformed in the sense of subsection \ref{subsec:KendalShapeSpace}. In other words, in Bayesian language we are giving empirical  prior on a neighbourhood of  the classical point estimates of $(c, b)=(\Bar{c},\bar{b})$ (see subsection \ref{subsec:Calssicalmethod}). In this new empirical prior choice the convergence of all the parameters will follow.
 
 The whole object data can be represented  as 3-dimensional array $\{w_{ijk}\}_{\tiny{\makecell{1\leq i\leq p\\1\leq j\leq 2\\1 \leq k \leq n}}}$.

 For $k=1,2, \cdots, n$ suppose $\bw_k=\{w_{1,k}, w_{2,k}, \cdots, w_{p,k}\}$, $\bz=\{z_1, z_2, \cdots, z_p\}$, where for all $i =1,2, \cdots,p$ each $w_i=(w_{i1}, w_{i2})^T$ and $z_i=(z_{i1}, z_{i2})^T$. We can write the bayesian regression  \eqref{complex_regression_1} as 
 \begin{eqnarray}
 \label{Model:2D_Regression}
 \begin{pmatrix}
 w_{i1k}\\w_{i2k}
 \end{pmatrix}=
 \begin{bmatrix}
 c_{1k}\\c_{2k}
 \end{bmatrix} +b_k\cdot \begin{bmatrix}
 \cos \theta_k &\sin\theta_k \\ -\sin\theta_k & \cos\theta_k
 \end{bmatrix} \begin{pmatrix}
 z_{i1k}\\z_{i2k}
 \end{pmatrix}+\begin{pmatrix}
 \epsilon_{i1k}\\\epsilon_{i2k}
 \end{pmatrix}
 \end{eqnarray}
 Assuming the objects coming from same family (e.g., same species or same  genera), the same error parameter $\sigma$ for all the landmark data points are justified on the basis of isotropic error variance (simplest model).  model \ref{Model:2D_Regression}  with bivariate Gaussian density for $f_{\bz}(\bw|\cdot)$ we can rewrite the regression model, for $k\in\{1, 2, \cdots, n\}$
 \begin{eqnarray}
 \label{Model:2D_Normal}
 \begin{pmatrix}
 w_{i1k}\\w_{i2k}
 \end{pmatrix}\sim \bN\left(
 \begin{bmatrix}
 c_{1k}\\c_{2k}
 \end{bmatrix} +b_k\cdot \begin{bmatrix}
 \cos \theta_k &\sin\theta_k \\ -\sin\theta_k & \cos\theta_k
 \end{bmatrix} \begin{pmatrix}
 z_{i1k}\\z_{i2k}
 \end{pmatrix}, \sigma^2 \cdot \begin{pmatrix}
1 & 0\\0 & 1
 \end{pmatrix} \right).
 \end{eqnarray}
 We asume the following prior for the parameter set $\lambda= (c_1, c_2, b, \theta, \sigma)$
 \begin{eqnarray}
 \label{2D_Prior}
\begin{bmatrix}
c_1\\c_2
\end{bmatrix}\sim\bN\left( \begin{bmatrix}
0\\0
\end{bmatrix}, \sigma^2 \begin{bmatrix}
1 &0\\0 & 1
\end{bmatrix}\right); \ b\sim \bN(0, \tau_{b}^{-1}); \theta \sim unif(-\pi,\pi), \ \sigma \sim unif(0,\infty). \end{eqnarray}
The  algorithm for such model \eqref{Model:2D_Normal} is stated in Algorithm \ref{Algorithm:MCMC_iso}.


 \begin{algorithm}[H]
  	\caption{Markov Chain Monte Carlo based-Sampling for BFPF (Bayesian Full Procruste Fit) for Isotropic Landmark Error Variance }\label{Algorithm:MCMC_iso}
 	\begin{algorithmic}[1]

 		\Require Object containing 2D-Landmark data $(\bz, \bw)$, Nlandmark, tune, Nsample
 		\Ensure A $(5\times Nsample)$ dimensional large matrix containing samples from posteriors of the parameters  $\lambda=(c=(c_1,c_2), b, \theta, \sigma)$. 
 	\Procedure{ MCMC Sampling}{(A version of Metropolis Sampling)}  
 	\State Initialize values of $\lambda=(c=(c_1,c_2), b, \theta, \sigma) =(c=(c_1(0),c_2(0)), b(0), \theta(0), \sigma(0))$    \Comment{May be chosen arbitrarily }

 	\Function{fratio }{$\lambda_1,\lambda_2$}
 	
 	\State Compute   log-density of (Likelihood $\times$ prior) assuming parameter values $\lambda_1=(c=(c_1,c_2), b, \theta, \sigma)_1$. \State Store in $f_1$; 
 	\State Compute log-density of (Likelihood $\times$ prior) assuming parameter values $\lambda_2=(c=(c_1,c_2), b, \theta, \sigma)_2$.
 	\State Store in $f_2$.
 	
 	\Return $\dfrac{f_1}{f_2}$
 	\EndFunction

 	\Function{purturb}{$\lambda$}
 		
 	\State For each given value of $\lambda \rightarrow\lambda_{old}$, choose a new set of parameter $\lambda_{new}$ inside a small-neighbourhood 
 		
 	\State (determined by tuning value "tune") of $\lambda_{old}$.
 	\State Generate $u\sim unif(0,1)$
 	\If{$u\leq 0.5$}
 	\State Select $\lambda \leftarrow  \lambda_{new}$
		\Else 
  		\State Select $\lambda \leftarrow  \lambda_{old}$
 		\EndIf
 		\EndFunction
 		 		
 	\Function{step}{$\lambda$, purturb}
 		
 		\State Pick new point
 		\State $\lambda_p = purturb(\lambda)$
 		 		
 		\State Compute Acceptance probability
 	 \State	$A \leftarrow   min (1, fratio(\lambda_p ,\lambda ))$
 		 	\State Accept new point with probability $A$.	
 		\EndFunction
 		\Function{run}{$\lambda, purturb, nsteps$}

 			\State Allocate matrix $res$
 			\For ($i=1, i \leq \ Nsample, \ i++$ )
 			
 				$res[i,] \leftarrow \lambda \leftarrow step(\lambda, purturb)$
 				\EndFor
 				
 		 		\Return $res=\{( c_1(t), c_2(t), b(t), \theta(t), \sigma(t))\}$ for all $t\in \{10001, 10002, \cdots, 20000\}$.
 		\EndFunction

 		\EndProcedure
 		 	\end{algorithmic}
 \end{algorithm}

 
 \begin{theorem}[Asymptotic Efficiency of BFPF]
 \label{theorem:Asymptotic Convergence Isotropic}
 Suppose  $\lambda_0=(b_0, c_0, \theta_0, \sigma_0)$ be the  truth value of the parameter  $\widetilde{\lambda}=(\widetilde{b}, \widetilde{c}, \widetilde{\theta}, \widetilde{\sigma})$. Consider the reparametrization $\eta(\lambda):=c\mathbf{1}_p+ bR_{\theta}\Bz$, with the truth value $\eta_0:= \eta(\lambda_0)=c_{0}\mathbf{1}_p+ b_0R_{\theta_0}\Bz$.  Then, under square error risk ($MSE_B$ of \eqref{Bayesian_criteria}), and for large samples (large values of number of objects $n$), the posterior density of $\widetilde{\lambda}\equiv \widetilde{\lambda}(n)$ is approximately normal. In other words, for large $n,p$,
 \begin{eqnarray}
 \label{Asymptotic Convergence Isotropic}
 \sqrt{n}\left(\bw_{BFPF} - \left( c^{T}_{0}\mathbf{1}_p+ b_0R_{\theta_0}\Bz\right)\right) \xrightarrow{d} N\left(0 , \frac{1}{(J^{-1})^TI(\lambda_0)(J^{-1})}\right),
  \end{eqnarray}
 where where $I(\lambda_0)$ is the fisher information of $\lambda_0$and $J$ is the Jacobian matrix such that  the $(i,j)$ th element of the Jacobian matrix $J$ is defined by
 $J_{ij} = \frac{\partial \eta_i}{\partial \lambda_j}$.
 \end{theorem}

\subsection{Problem of Ill-Posedness  \& Remedy using Empirical Prior}
\label{subsec:ill-posedness}

Assumption \ref{assumption:Isotropy} is needed to by pass the pathological scenario of having more unknown parameters than number of observations, known as ``ill-posed" problem. In fact, even under non-singularity assumption of variance-covariance matrix of landmarks, we may need additional assumption of indepency, otherwise the number of parameters involving variance-covariance matrix will be high. This type of assumptions may not always hold in practice, and under certain circumstances we may relax the assumption of Independency using empirical prior (a prior which depends on data). 

In fact, \cite{mardia2000statistical}proposed a fully multivariate test of directional asymmetry
for the case of object symmetry for bypassing  the assumption of equal, independent,
and isotropic variation at all landmarks.

The symmetry in morphological structures is  a big concern. It can cause serious statistical problems, for instance ill-conditioned covariance matrices if all the landmark configurations are very nearly symmetrical \citep{klingenberg2002shape}, \citep{bookstein1996combining}. Algebraically, symmetry (even if symmetry is not perfect) induces predictability which in tern makes
linear dependence among the landmarks, and therefore, the
covariance matrix of landmark positions will be singular (for imperfct symmetry,  covariance matrices will be ill-conditioned). This
causes difficulties for any statistical procedures that use
the inverse or determinant of the dispersion matrix, only remedy for such problem is  by taking the
symmetry of the forms into account explicitly and thereby adjusting
the analysis accordingly \citep{klingenberg2002shape}, the central idea of which revolves around  finding proper method for partitioning
the total shape variation of landmark configurations with object symmetry into components of symmetric variation
among individuals and asymmetry.

\section{Novel Measure of  Intra -species and Inter-species Shape Variability }
\label{sec:KLdivPosterior}

Suppose we have two sets of landmark based objects (3D array), namely $\bz_n$ and $\bw_n$. Our interest is to construct suitable statistical hypothesis test to infer  whether they belong to same taxa or different taxa, based on Procrustes variance among them. For null hypothesis of same taxa, we may think of the landmark data to come from  same distribution with same  posterior mean, posterior variance. Consider the registration object $v_{register}$ and $w_{register}$.  For $i \in\{0,1\}$, let $\theta_i$ be model parameter for model $M_i$. For instance, $\theta_0=\Sigma$ and $\theta_i=(\lambda_i,\lambda_2, \Sigma_1, \Sigma_2$ for the simplest case with Gaussian likelihood. Then,
\begin{eqnarray}
H_0: \quad &\bz_n \sim \bw_n\sim   \bN \left((z_{register}, w_{register})^T, \Sigma \right);\\
 M_0: \quad & L_{n}(\theta_0|\bz_n,\bw_n, M_0) =\bN \left((z_{register}, w_{register})^T, \begin{bmatrix}
 \Sigma &0 \\0 & \Sigma
 \end{bmatrix} \right).
\end{eqnarray}
For alternate hypothesis of different taxa, we may think of the landmark data to come from  mixture-distribution with different posterior mean, posterior variance.
\begin{eqnarray}
H_1: \quad & \bz_n|\lambda_1 \sim \bN (z_{register}, \Sigma_1) ; \\
 & \bw_n |\lambda_2\sim  \bN (w_{register}, \Sigma_2);,  \\
 M_1: \quad &L_{n}(\theta_1|\bz_n,\bw_n, M_1) = \lambda_1 \bN (z_{register}, \Sigma_1) + \lambda_2 \bN (w_{register}, \Sigma_2).
\end{eqnarray}
Then the Bayes factor (interpreted as the quantification of the evidence of model $M_0$ against model
$M_1$, given objects $\bz_n, \bw_n$) of model $M_0$ against $M_1$ is given by
\begin{eqnarray}
\label{bayes_factor}
B_{n}^{12}=\dfrac{m\left( \bz_n,\bw_n|M_0 \right)}{m\left(\bz_n,\bw_n|M_1 \right)},
\end{eqnarray}
where, for $i\in\{0,1\}$, marginal densities for the two models be $m\left( \bz_n,\bw_n|M_i \right)= \int_{\theta_i} L_{n}(\theta_i|\bz_n,\bw_n, M_i) \ \pi(d\theta_i|M_i)$ respectively, with prior $\pi(\theta_i|M_i)$. \cite{chatterjee2020short} theoretically proved that, asymptotically bayes factor goes to a version of Kulberk-Leibler divergence of densities of two competing models, even under miss-specifications. Formally, 
\begin{eqnarray}
\lim_{n\to \infty} \log \left( B_{n}^{12} \right)= h_1(\theta_1)-h_0(\theta_0), 
\end{eqnarray}
where 
\begin{eqnarray}
h_{i}(\theta_i)=\lim_{n\to\infty}\dfrac{1}{n} \bE\left(\log\left\lbrace\dfrac{m(\bz_n,\bw_n|M_i)}{L_{n}(\theta_i|\bz_n,\bw_n,M_i)}\right\rbrace  \right).
\end{eqnarray}
Under Gaussian distribution assumption, one of the main application of Mahalanobis distance \citep{mahalanobis1936generalized} is to measure the distance between two densities which takes account of intra and inter-species (or, inter-genera) variability. This is because expression of   Mahalanobis distance for multivariate normal is very similar to  Kullback–Leibler divergence, a measure of difference from two distributions. After estimating the mean and variance parameter for landmark data of  two hypothesized species populations, we can compute the Kullback–Leibler divergence from $\bN_2(\boldsymbol\mu_2, \boldsymbol\Sigma_2)$  to $\bN_1(\boldsymbol\mu_1, \boldsymbol\Sigma_1)$, for non-singular matrices $\Sigma_2$ and $\Sigma_1$, is:

\begin{eqnarray}
\label{KL_div}
D_\text{KL}(\bN_2 \| \bN_1) = { \dfrac{1}{2} } \left\{ \operatorname{tr} \left( \boldsymbol\Sigma_1^{-1} \boldsymbol\Sigma_2 \right) + \left( \boldsymbol\mu_1 - \boldsymbol\mu_2\right)^{\rm T} \boldsymbol\Sigma_1^{-1} ( \boldsymbol\mu_1 - \boldsymbol\mu_2 ) - d +\ln { \dfrac{|  \boldsymbol \Sigma_1 |} {| \boldsymbol\Sigma_2 |} } \right\},
\end{eqnarray}

where $d$ is the dimension of the vector space, for instance $d=2$ in 2D landmark data. From BFPF methods discussed so far, we can get estimates  $\widetilde{\boldsymbol\mu_1} , \widetilde{\boldsymbol\mu_2}, \widetilde{\boldsymbol\Sigma_1}, \widetilde{ \boldsymbol\Sigma_2}$ and hence, from \eqref{KL_div} we can get an estimate of the divergence between two populations (characterized by densities $\bN_1(\cdot)$ and $\bN_2(\cdot)$).  We may put species variability problem in the statistical hypothesis frame:
\begin{eqnarray}
\label{hypothesis_KL_div} 
\bH_0: \ \ D_\text{KL}(\bN_2 \| \bN_1)  < \epsilon; \\
\bH_1: \ \ D_\text{KL}(\bN_2 \| \bN_1)  \geq \epsilon; 
\end{eqnarray}
for some threshold $\epsilon >0$.  The sample-version of KL-divergence $\widetilde{D_\text{KL}}(\bN_2 \| \bN_1)$ will form a discrepency statistic, based on which we can calculate Bayesian predictive p-value for inference (see Appendix \ref{Appendix}).


\section{Applications}
\label{sec:Application}
\subsection{Convergence diagnostics using simulated random convex-concave polygon in 2D}

\begin{figure}[H]
  \centering
  \includegraphics[width=0.9\linewidth]{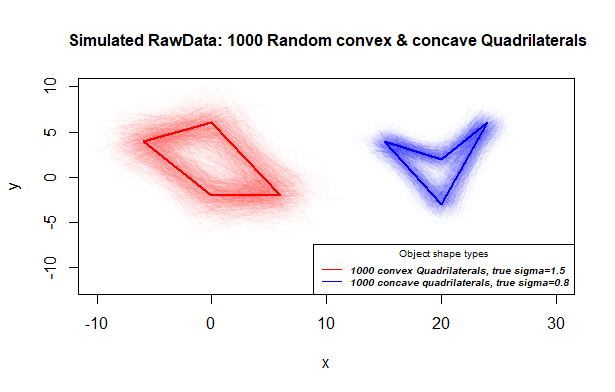}
    \label{fig:rawquadrilateral}
    \caption{  Plot of actual simulated 1000 convex quadrilateral raw-objects (red) with isotropic landmark variance  $\sigma=1.5$ and concave quadrilateral raw-objects (blue) with isotropic landmark variance  $\sigma=0.8$.}
\end{figure}

We generate 1000 quadrilateral-objects  (set of 4 landmarks for a single quadrilateral),  each from two  different types of shapes with truth values of variance parameter $(\sigma)$ as follows.
\begin{description}
    \item [Convex shaped Quadrilateral:] isotropic  landmark variance with $\sigma_0=1.5$,
     \item [Concave shaped Quadrilateral:] isotropic  landmark variance with $\sigma_0=0.8$.
\end{description}
Figure \ref{fig:rawquadrilateral} shows the raw-object plot.
\paragraph{Objectives:} As we already know the truth value of parameter(s) $\sigma_v=1.5$ and $\sigma_c=0.8$, we wish to test the consistency of our Bayesian models computationally. here the assumption of isotropic error variance  and assumption \ref{Ass_Intuition-1} are both  automatically  satisfied because we are simulating deliberately using isotropic variance for the corners (landmarks) of those quadrilaterals.

First, we transform the objects into pre-shape space.  Then We assume the following model for both convex and concave quadrilaterals except different variance parameters $\sigma_c$ for concave quadrilateral objects and $\sigma_v$ for convex quadrilateral objects):
\begin{eqnarray}
 \label{model_for quadrilaterals}
\begin{bmatrix}
vq_{i,1,k}\\ vq_{i,2,k}
\end{bmatrix}\sim\bN\left( \begin{bmatrix}
0\\0
\end{bmatrix}, \sigma_{v}^{2} \begin{bmatrix}
1 &0\\0 & 1
\end{bmatrix}\right); \quad \begin{bmatrix}
cq_{i,1,k}\\cq_{i,2,k}
\end{bmatrix}\sim\bN\left( \begin{bmatrix}
0\\0
\end{bmatrix}, \sigma_{c}^{2} \begin{bmatrix}
1 &0\\0 & 1
\end{bmatrix}\right).
\end{eqnarray}

First, we assume non-informative  prior for the parameter set $\sigma_{v}\sim \sim unif(0,\infty), \sigma_{c}\sim unif(0,\infty)$. As we can observe, we can achieve convergence to the truth value, 
 
\begin{figure}[H]
   \centering
  \centering
  \includegraphics[width=0.9\linewidth]{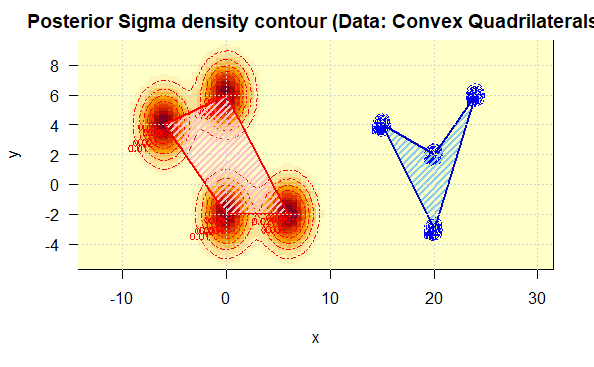}
 \label{fig:contourquadrilateral}
    \caption{  Plot of posterior density contour   of isotropic landmark variance. Compare with figure \ref{fig:rawquadrilateral} with truth  $\sigma=1.5$ for convex quadrilaterals (red)  and $\sigma=0.8$ for concave quadrilaterals (blue). The assumption of  isotropic landmark variance  is crucial here.}
\end{figure}


\subsection{Bayesian Procrustes Analysis on ``apes'' Data: fluctuation in face-shape for Male vs. Female}
The data "apes" can be found in R package ``shapes". It is taken from \cite{o1993sexual} and also in the famous book of statistical shape analysis \citep{dryden2016statistical}. The data falls in the category of geometric morphometrics landmark-based data. It has the following attributes.
\begin{description}
    \item[apes\$x] : An array of dimension $8 \times  2 \times 167$,
\item[apes \$group:] Species and sex of each specimen:
\begin{enumerate}
    \item ["gorf":] Female gorilla skull data. 8 landmarks in 2 dimensions, 30 individuals,
    \item ["gorm":] Male gorilla skull data. 8 landmarks in 2 dimensions, 29 individuals,
    \item  ["panf":] Female chimpanzee skull data. 8 landmarks in 2 dimensions, 26 individuals,
    \item ["panm":] Male chimpanzee skull data. 8 landmarks in 2 dimensions, 28 individuals,
    \item ["pongof":] Female orang utan skull data. 8 landmarks in 2 dimensions, 30 individuals,
    \item ["pongom":] Male orang utan skull data. 8 landmarks in 2 dimensions, 30 individuals.
\end{enumerate}

\end{description}

\paragraph{Objective:} To obtain Bayesian evidence for the hypothesis: ``male primate manifests more fluctuation in face-shape than females".

\begin{figure}[H]
\centering
\begin{minipage}{.48\textwidth}
  \centering
  \includegraphics[width=\linewidth]{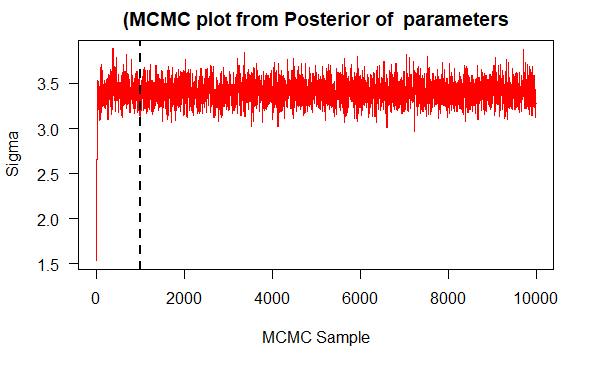}
  \caption{MCMC mixing plot of face-shape landmark data of Gorilla-female. Figure shows satisfactory mixing for $10000$ samples with $1000$ burn-ins with tune=$0.05$, even from arbitrary start $=1.5$}
  \label{fig:mcmcgorillafemale}
\end{minipage}%
\hspace{0.1cm}
\begin{minipage}{.48\textwidth}
  \centering
  \includegraphics[width=\linewidth]{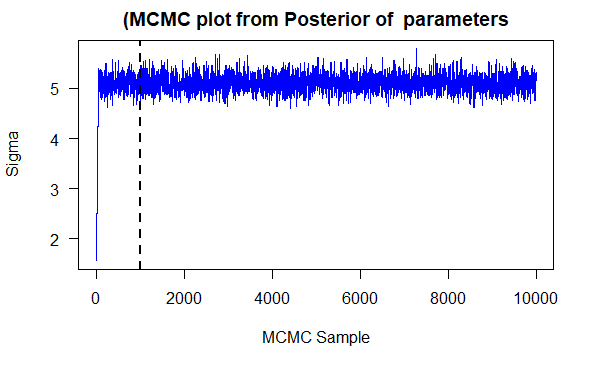}
  \caption{MCMC mixing plot of face-shape landmark data of Gorilla-male. Figure shows satisfactory mixing for $10000$ samples with $1000$ burn-ins with tune=$0.05$, even from arbitrary start $=1.5$}
  \label{fig:mcmcgorillamale}
\end{minipage}
\end{figure}

\begin{figure}[H]
  \centering
  \includegraphics[width=0.9\linewidth]{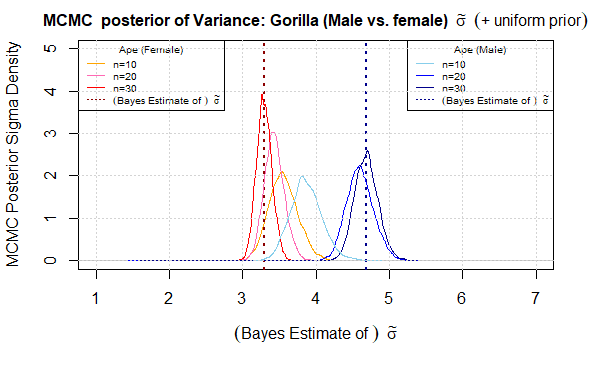}
    \label{fig:GorillaMaleVs.Female}
    \caption{  Plot of MCMC posterior of Gorilla male vs. female Variance in Landmark after Pre-shape Space-formation}
\end{figure}

\begin{figure}[H]
  \centering
  \includegraphics[width=0.9\linewidth]{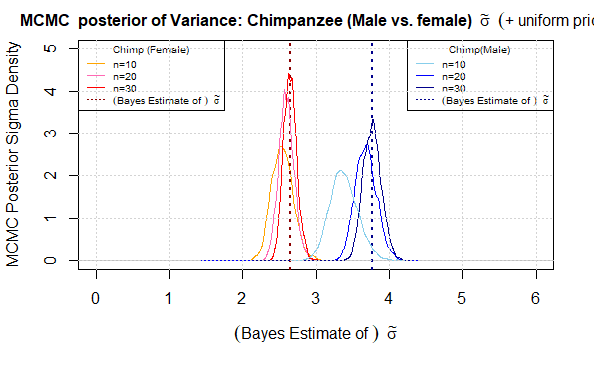}
    \label{fig:ChimpanzeeMaleVs.Female}
    \caption{  Plot of MCMC posterior of Chimpanzee male vs. female Variance in Landmark after Pre-shape Space-formation tune=$0.05$, even from far start $=1.5$}
\end{figure}

\begin{figure}[H]
  \centering
  \includegraphics[width=0.9\linewidth]{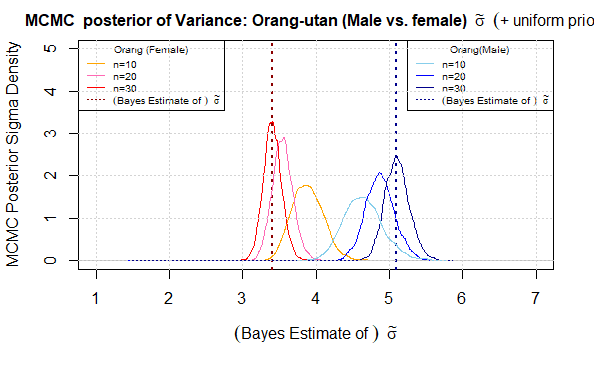}
    \label{fig:OrangMaleVs.Female}
    \caption{  Plot of MCMC posterior of Orang-utan male vs. female Variance in Landmark after Pre-shape Space-formation}
\end{figure}

\paragraph{Conclusion from Bayesian Procrustes Analysis using novel R package \texttt{BPviGM1}}
It is being observed from posterior Face-shape  variance density  comparison-plot that there are Bayesian evidence for more face-shape variability in  all 3 primates (Gorilla, Chimpanzee \& Orang-utan) male than the same for respective  Females. In other words, our result supports the hypothesis that primate male-face  may be genetically viable to more shape-variation than the same for females.

\subsection{Over-fit Problem using Bayesian Procrustes Analysis  on Objects already kept in Shape-Space using classical Method}\label{subsec:overfit}
\begin{figure}[H]
  \centering
  \includegraphics[width=0.9\linewidth]{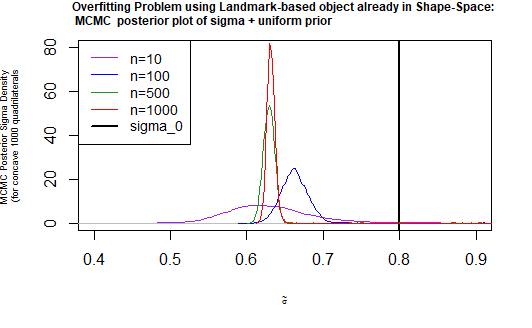}
    \label{fig:overfit problem}
    \caption{  Plot of MCMC posterior of  simulated 1000 concave quadrilateral raw-objects  with isotropic landmark variance  $\sigma=0.8$, demonstrating the problem of over-fit}
\end{figure}

For Bayesian Procrustes analysis, we may use landmark data in configuration space, or in pre-shape space. Problem arises when we try to make Bayesian inference on  landmark already in shape-space by use of classical methods. This is because of over-fitting arising from twice use of same data. This results in apparent decrease of variance. For instance, although we know the actual variance, in fig \ref{fig:overfit problem} we demonstrate ``over-fit" problem, where the posterior of variance fails to capture the truth (black line) and demonstrates lesser variation than it actually is. Here we have used landmarks already from shape-space for concave quadrilateral data.

\section{ Novel R Package \texttt{BPviGM1} \& Source Code Availability  }
\label{sec:RpackageSourcecode}
The computation part for the implementation of the Bayesian method and posterior sampling is a challenging task. Moreover,  we are not aware of any direct Bayesian R package for Procrustes analysis using Geometric Morphometric objects.  Hence we have built a novel and simple R package \texttt{BPviGM1} (Bayesian   Procrustes Variance-based Inferences  in Geometric Morphometrics) which essentially contains the R code version of the Bayesian methods discussed in this paper.

\subsection{ Novel  R package \texttt{BPviGM1}}
A novel R package \texttt{BPviGM1}  ("Bayesian Procruste Analysis using  Landmark based on Isometric/Anisometric Error under Geometric Morphometrics"), which mainly  contains functions corresponding to  R code implimentation of algorithm for MCMC based posterior sampling. These functions correspond to the above-discussed novel bayesian Procruste Analysis using isometric/anisometric Landmark error,   are available as a novel R package  through GitHub \url{https://rdrr.io/github/debashischatterjee111/BPviGM1/},  (\url{https://github.com/debashischatterjee111/BPviGM1}).

\subsection{Installation of R package \texttt{BPviGM1}}
 The package is currently available freely from Github. If download from GitHub,  you can use devtools by the commands:

\begin{rc}
   install.packages("devtools")
   require(devtools)
   install_github("debashischatterjee111/BPviGM1")
\end{rc}


Alternatively,  you first install  R package "githubinstall", thereby call \texttt{BPviGM1} from it, using following command in R:

\begin{rc}
  install.packages("githubinstall")
  require(githubinstall)
  githubinstall("BPviGM1")
   \end{rc}

Once the packages are installed, it needs to be made accessible to the current R session by the commands:

\begin{rc}
   require(BPviGM1)
\end{rc}



\subsection{Main functions described in Novel R package \texttt{BPviGM1}}
There will be mainly 4 kinds of functions available in the R package, all of which generates samples from posterior of the bayesian regression parameters discussed  in section \ref{sec:NovelBayesian}. For details, type R command

\begin{rc}
  ? <function>
   \end{rc}



\begin{description}
	
		\item[\myV{\texttt{Cmat}\{BPviGM1\}}:] This function changes a 3D array to a matrix using row-bind.
		
	\item[\myV{\texttt{Helmert}\{BPviGM1\}}:] "Helmert" computes The Helmert sub-matrix,
				
\item[\myV{\texttt{MCMCpostPsample2D}\{BPviGM1\}}:]MCMC posterior sampling for 2D landmark data (in Pre-shape space) (Gaussian likelihood with Isotropic Error Variance)

	\item[\myV{\texttt{MCMCpostsample2D}\{BPviGM1\}}:] MCMC posterior sampling for 2D landmark data (Gaussian likelihood with Isotropic Error Variance), able to draw posterior from 5 parameters "c1","c2", "b", "theta", "Sigma".
			
\item[\myV{\texttt{PLOTpostvar2D}\{BPviGM1\}}:] "PLOTpostvar2D" Plot of posterior of Landmark variance parameter from MCMC sampling.
	
	\item[\myV{\texttt{PPLOTpostvar2D}\{BPviGM1\}}:] "PPLOTpostvar2D" Plot of posterior of Landmark variance parameter from MCMC sampling ( single or double parameters)

\item[\myV{\texttt{Pfratio2D}\{BPviGM1\}}:] 2D landmark data in Pre-shape space (Gaussian likelihood with Isotropic Error Variance) as the name suggest, it evaluates fratio for two parameter vectors.

		\item[\myV{\texttt{fratio2D}\{BPviGM1\}}:] 2D landmark data(Gaussian likelihood with Isotropic Error Variance) as the name suggest, it evaluates fratio for two parameter vector (for multi-dimensional vector)

	\item[\myV{\texttt{purturb2D}\{BPviGM1\}}:] 2D landmark data(Gaussian likelihood with Isotropic Error Variance) generates pertrubed point from 5 parameter space.
		\item[\myV{\texttt{Ppurturb2D}\{BPviGM1\}}:] 2D landmark data(Gaussian likelihood with Isotropic Error Variance) generates purtubed point from 5 parameter space (single or double parameter set).

		\item[\myV{\texttt{TMCMCpostsample2D}\{BPviGM1\}}:] MCMC posterior sampling for 2D landmark data (Gaussian likelihood with Isotropic Error Variance), able to draw posterior from 5 parameters "c1","c2", "b", "theta", "Sigma".

		\item[\myV{\texttt{Prun2D}\{BPviGM1\}}:]  2D landmark data(Gaussian likelihood with Isotropic Error Variance) Accepts new parameter vector point with probability alpha (single or double parameter set).

		\item[\myV{\texttt{step2D}\{BPviGM1\}}:] 2D landmark data(Gaussian likelihood with Isotropic Error Variance) Accepts new parameter 5*1 point with probability alpha.

		\item[\myV{\texttt{Pstep2D}\{BPviGM1\}}:] 2D landmark data(Gaussian likelihood with Isotropic Error Variance) Accepts new parameter 5*1 point with probability alpha (single or double parameter set).

			\item[\myV{\texttt{Simulated Polygon Dataset(s)}\{BPviGM1\}}:] ccq, cvq, fivetr.

					\item[\myV{\texttt{MCMCpostsample3D}\{BPviGM1\}}:] function for MCMC sampling for 2D landmark data (Gaussian likelihood with general anisotropic  Error Variance with Empirical Bayes Prior), \myR{**( To do in 2021)}
			\item[\myV{\texttt{PLOTpostvar3D}\{BPviGM1\}}:] function for MCMC sampling for 3D landmark data (Gaussian likelihood with general anisotropic  Error Variance with Empirical Bayes Prior). \myR{**( To do in 2021)}
				\item[\myV{\texttt{COMPAREpostvar2D}\{BPviGM1\}}:] function for MCMC sampling for 3D landmark data (Gaussian likelihood with general anisotropic  Error Variance with Empirical Bayes Prior). \myR{**( To do in 2021)}
				\item[\myV{\texttt{COMPAREpostvar3D}\{BPviGM1\}}:] function for MCMC sampling for 3D landmark data (Gaussian likelihood with general anisotropic  Error Variance with Empirical Bayes Prior). \myR{**( To do in 2021)}
				\end{description}


\paragraph{Detailed Examples of the usage of R functions in "BPviGM1":} See in Appendix \ref{Subsec:Appendix_R}.

	

\subsection{source code availability for Application part using R package "BPviGM1"}

The following information is to be noted regarding source code availability for Application part using R package \texttt{"BPviGM1"}.

\begin{description}
	
	\item[Source code:] All the source code of all the simulations and data analysis conducted in this paper are freely available  through GitHub \url{https://rdrr.io/github/debashischatterjee111/BPviGM1/},  (\url{https://github.com/debashischatterjee111/Sourcecode1}).
	
\end{description}

\section{Conclusion}\label{sec:Conclusion}
Bayesian approach to geometric morphometrics is a new  interdisciplinary branch with huge potential because of many natural advantages of bayesian methods that classical analysis lacks. In this paper, novel bayesian models with different kind of priors has been implemented for the analysis of morphometric variability along with test for model validation under different situations. Moreover, a novel R package has been  provided for that, with  explanation for how to use it. Please send suggestions and report bugs to \url{https://github.com/debashischatterjee111/BPviGM1/issues}, or email to \textit{cdebashis.r@gmail.com}.

\section*{Acknowledgments}

Author thanks Geological Studies Unit of Indian Statistical Institute for providing  motivations and encouragements  to pursue research on novel bayesian methods for Geometric morphometrics.    Author also  thanks the anonymous reviewers for their constructive comments and suggestions.


\appendix
\section{Appendix}
\label{Appendix}

\subsection{Proofs}

\subsubsection{Proof of Theorem \ref{theorem_1} }

\begin{Proof}
	Both theorems follows trivially from properties of gaussian density. For ddetailed discussion, we refer to \cite{mahalanobis1936generalized} and  \cite{mardia2000statistical}.
\end{Proof}

	 \subsubsection{Proof of Theorem \ref{theorem:Asymptotic Convergence Isotropic}}

	 	Here we stae two Lemmas (refer to \cite{lehmann2006theory} for detailed proof).
	 \begin{lem}\label{Lemm_posterior_Mean_Conv}
	 	suppose $\lambda_0$ be the truth value of unknown parameter $\lambda$, then under $MSE_B$, the posterior mean $\widetilde{\lambda}\equiv \widetilde{\lambda}(n)$ satisfies
	 	\begin{eqnarray}
	 	\label{posterior_Mean_Conv}
	 	\sqrt{n}(\widetilde{\lambda}(n) - \lambda_0) \to N\left(0 , \frac{1}{{\mathcal I}(\lambda_0)}\right),
	 	\end{eqnarray}
	 	in distribution, where $\mathcal I (\lambda_0)$ is the fisher information of $\lambda_0$.
	 	It follows that the Bayes estimator $\widetilde{\lambda}$ under MSE is asymptotically efficient.
	 \end{lem}
	 
	 \begin{lem}[Reparametrization of Fisher Information]
	 	\label{Lemm_reparametrization}
	 	Let $ \lambda$ and $\eta(\lambda):=c^{T}\mathbf{1}_p+ bR_{\theta}\Bz$ be  two  parametrizations of our Bayesian regression  estimation problem, where $\eta(\lambda)$ is  continuously differentiable function of $ \lambda$ with the  truth value $\eta_0:= \eta(\lambda_0):=c^{T}_{0}\mathbf{1}_p+ b_0R_{\theta_0}\Bz$.    then,
	 	\begin{eqnarray}
	 	\label{reparametrization_eqn}
	 	{\mathcal I}({\lambda}) = {\boldsymbol J}^\textsf{T} {\mathcal I} ({ \eta}({ \lambda})) {\boldsymbol J},
	 	\end{eqnarray}
	 	where $J$ is the Jacobian matrix such that  the $(i,j)$ th element of the Jacobian matrix $J$ is defined by
	 	$J_{ij} = \frac{\partial \eta_i}{\partial \lambda_j}.$ The detailed proof is given \cite{lehmann2006theory}.
	 \end{lem}

	\begin{Proof}[Proof of Theorem \ref{theorem:Asymptotic Convergence Isotropic}]
		
		Directly follows from Lemma \ref{Lemm_reparametrization} and Lemma \ref{Lemm_posterior_Mean_Conv}. 
	\end{Proof}


\subsection{ Helmert matrix}\label{subsec:Helmert}
Standard Helmert matrix \citep{helmert1876genauigkeit, lancaster1965helmert}  of order $n$ is an orthogonal square matrix such that
\begin{enumerate}
    \item[(i)] $h_{ij}=0$ for $j>i>1$,
    \item[(ii)]  $h_{1j}=+\sqrt{w_{j}}; \ \ w_{j}>0; \ \ \displaystyle\sum_{i=1}^{n} w_j =1$ (in the strict sense).
\end{enumerate}
 Helmert matrix in the strict sense has been used widely in statistics. A generalized helmert matrix is which  can be transformed by permutations of its rows and
columns or by transposition or by change of sign of rows, to a form of standard Helmert matrix \citep{helmert1876genauigkeit, lancaster1965helmert}.

 \begin{theorem}[\cite{helmert1876genauigkeit, lancaster1965helmert}]
 All standard Helmert matrix of size $n$ has rank $(n-1)$. In other words, standard Helmert matrix of size $n$   depends on $(n-1)$ independent parameters. 
 \end{theorem}
 
 \begin{rmk}[\cite{lancaster1965helmert}]
    As  standard Helmert matrix of size $n$ is orthogonal, we can take the $(n-1)$ independent parameters of a standard Helmert matrix as the angles of certain rotations, namely in the plane of the $1st$ and $j-th$ coordinate axes, for $j=2, 3,\cdots, n$. Moreover,  A Helmert matrix comes in the evaluation of the Jacobian of the transformation from rectangular Cartesian to polar coordinates and vice versa.
    \end{rmk}
    
\subsubsection*{Examples of Helmert matrix}
\paragraph{Example 1:} \cite{helmert1876genauigkeit} has taken  $h_{1j}=\sqrt{w_{j}}=\dfrac{1}{\sqrt{n}}$.

\paragraph{Example 2: (Rotation matrices in 2D)}  For $n = 2$, there cannot  be any zero above the diagonal and below the first row. Hence, the 2D rotation matrix $R_{\theta}$ are standard Helmert matrix, where
\begin{eqnarray}
\label{helmert_2D_rotation}
R_{\theta}=\begin{bmatrix}
 \cos \theta &\sin\theta \\ -\sin\theta & \cos\theta
 \end{bmatrix} 
\end{eqnarray}

Rotation matrices on euclidean space are square matrices,  characterized by orthogonal matrices with determinant 1. A square matrix $R$ is a rotation matrix if and only if $R^T = R^{-1}$ and $determinant(R) = 1$. The set of all orthogonal matrices of size n with determinant $\pm 1$ forms the orthogonal group $O(n)$, and he set of all orthogonal matrices of dimension $d\times d$  with determinant $1$ forms the special orthogonal group $SO(n)$, for example,  the rotation group $SO(2)$ in 2D and the rotation group $SO(3)$ in 3D. 

Let $L= \ = \ \left \lbrace \begin{pmatrix}a & b \\ c & -a \end{pmatrix}: a^2 + bc + 1 = 0 \right\rbrace.$ Let $I$ be the identity matrix, then it can be shown that  $P_m = \{ xI + yt : x, y \in \Re, t\in L \}$ is a plane of matrices isomorphic to complex plane $\C$. It follows  from Eular's formula that, any complex rotation 
$ \exp (\theta t) \ = \ \cos(\theta) I + t \sin(\theta)$  has a one-one correspondence with rotation matrix.

In 2D, we may use matrix notation for counterclockwise rotation by angle $\theta$ for orthogonal matrix multiplication (rotation matrix), for example
\begin{eqnarray}\label{rotation matrix 2D}
\begin{bmatrix}
\cos\theta & -\sin\theta \\
\sin\theta & \cos\theta
\end{bmatrix}
\begin{bmatrix}
x \\
y
\end{bmatrix}= 
\begin{bmatrix}
x\cos\theta  -y\sin\theta \\
x\sin\theta + \cos\theta
\end{bmatrix}
\end{eqnarray}

\paragraph{Example 3: (Rotation matrices in 3D)}

\begin{eqnarray}
 \mathbf{R}_{\theta_X} = \begin{bmatrix} 1 & 0 & 0\\ 0 & \cos \theta_x  & -\sin \theta_x \\ 0 & \sin \theta_x  & \cos \theta_x  \end{bmatrix};\quad
\mathbf{R}_{\theta_Y}  = \begin{bmatrix} \cos \theta_y & 0 & \sin \theta_y\\ 0 & 1 & 0\\ -\sin \theta_y & 0 & \cos \theta_y \end{bmatrix};\quad
 \mathbf{R}_{\theta_Z} = \begin{bmatrix} \cos \theta_z  & -\sin \theta_z  & 0\\ \sin \theta_z  & \cos \theta_z  & 0\\ 0 & 0 & 1 \end{bmatrix}
\end{eqnarray}

\subsection{Bayesian Predictive p-value}
We use Bayesian p-values when one would like to check how a model fits the data. Given a model M, we wish to examine how well it fits the observed data $x_{obs}$ based on a statistic $T$. In other words, it measures the goodness of fit of data and model. Consider model $M$ with probability density function $f(x|\theta)$ and with prior $g(\theta)$.  The prior predictive p-value  under the predictive distribution  is
\begin{eqnarray}\label{pred1}
p=P(T(x)\geq T(x_{obs})|M)=\int T(x)\geq T(x_{obs}) h(x) \ dx,
\end{eqnarray}

where $h(x)dx=\int f(x|\theta)g(\theta) \ d\theta$ is the prior predictive density.
	\eqref{pred1} may be influenced by the choice of the prior \citep{ghosh2007introduction}. For this reason, the posterior predictive p-value was introduced. If we consider empirical prior which depends on the observed data $g( \theta |x_{obs})$, then, $h(x|x_{obs})=\int f(x|\theta)g(\theta|x_{obs}) \ d\theta$.



\subsection{Gibbs-Sampling Algorithm for Bayesian  Procrustes Analysis under fully-known conditionals}

\begin{algorithm}\label{Algorithm_Gibbs}
	\caption{Gibbs Sampling for BFPF (Bayesian Full Procruste Fit), When all Conditional distributions are known]}
	\begin{algorithmic}[1]
				\Procedure{Gibbs Sampling}{from posterior of $(b, c, \theta, \sigma)$}       
		\State Initialize values of $(b, c, \theta, \sigma)=(b(1), c(1), \theta(1), \sigma(1))$    \Comment{May be chosen arbitrarily}
		\State Assign step $t=1$
		
		\While{ step $t \leq 1000+20000$}  \Comment{Assuming burn in value to be 1000}
		\State Sample $c(t+1)$  from $\pi\left(c(t)| b(t) , \theta(t) , \sigma(t), \bz, \bw \right)$
		\State Sample $b(t+1)$  from  $\pi\left(b(t)| c(t) , \theta(t) , \sigma(t), \bz, \bw \right)$,
		\State Sample $\theta(t+1)$  from $ \pi\left(\theta(t)| c(t) , b(t) , \sigma(t), \bz, \bw \right)$,
		\State Sample $\sigma(t+1)$  from $\pi\left(\sigma(t)| b(t) , \theta(t) , \bz, \bw \right)$.
		\State  $t \leftarrow t+1$
		\EndWhile  \label{Sampling Loop}
		
		\Return $\{(b(t), c(t), \theta(t), \sigma(t))\}$ for all $t\in \{10001, 10002, \cdots, 20000\}$.
		
		\EndProcedure
	\end{algorithmic}
\end{algorithm}




\subsection{Detailed Examples of usage of R functions in ``BPviGM1"}
\label{Subsec:Appendix_R}

Refer to R package ``BPviGM1" R Suppliment(s) kept in Github: \url{https://github.com/debashischatterjee111/Sourcecode1}.


\bibliographystyle{authordate1}  
\bibliography{bibBPvI}  

\begin{thebibliography}{}

\bibitem[\protect\citename{Bookstein, }1996]{bookstein1996combining}
Bookstein, Fred~L. 1996.
\newblock Combining the tools of geometric morphometrics.
\newblock {\em Pages  131--151 of:} {\em Advances in morphometrics}.
\newblock Springer.

\bibitem[\protect\citename{Bookstein, }1997]{bookstein1997morphometric}
Bookstein, Fred~L. 1997.
\newblock {\em Morphometric tools for landmark data: geometry and biology}.
\newblock Cambridge University Press.

\bibitem[\protect\citename{Chatterjee {\em et~al.}, }2020]{chatterjee2020short}
Chatterjee, Debashis, Maitra, Trisha, \& Bhattacharya, Sourabh. 2020.
\newblock A short note on almost sure convergence of Bayes factors in the
  general set-up.
\newblock {\em The American Statistician}, {\bf 74}(1), 17--20.

\bibitem[\protect\citename{Dem{\v{s}}ar {\em et~al.},
  }2020]{demvsar2020bayes4psy}
Dem{\v{s}}ar, Jure, Repov{\v{s}}, Grega, \& {\v{S}}trumbelj, Erik. 2020.
\newblock bayes4psy—An Open Source R Package for Bayesian Statistics in
  Psychology.
\newblock {\em Frontiers in Psychology}, {\bf 11}.

\bibitem[\protect\citename{Dryden \& Mardia, }1998]{dryden1998statistical}
Dryden, Ian~L, \& Mardia, Kanti~V. 1998.
\newblock {\em Statistical shape analysis: Wiley series in probability and
  statistics}.

\bibitem[\protect\citename{Dryden \& Mardia, }2016]{dryden2016statistical}
Dryden, Ian~L, \& Mardia, Kanti~V. 2016.
\newblock {\em Statistical shape analysis: with applications in R}.
\newblock  Vol. 995.
\newblock John Wiley \& Sons.

\bibitem[\protect\citename{Fox {\em et~al.}, }2016]{fox2016applications}
Fox, Neil~I, Micheas, Athanasios~C, \& Peng, Yuqiang. 2016.
\newblock Applications of Bayesian Procrustes shape analysis to ensemble radar
  reflectivity nowcast verification.
\newblock {\em Atmospheric Research}, {\bf 176}, 75--86.

\bibitem[\protect\citename{Ghosh {\em et~al.}, }2007]{ghosh2007introduction}
Ghosh, Jayanta~K, Delampady, Mohan, \& Samanta, Tapas. 2007.
\newblock {\em An introduction to Bayesian analysis: theory and methods}.
\newblock Springer Science \& Business Media.

\bibitem[\protect\citename{Goodman, }1963]{goodman1963statistical}
Goodman, Nathaniel~R. 1963.
\newblock Statistical analysis based on a certain multivariate complex Gaussian
  distribution (an introduction).
\newblock {\em The Annals of mathematical statistics}, {\bf 34}(1), 152--177.

\bibitem[\protect\citename{Gunz \& Mitteroecker, }2013]{gunz2013semilandmarks}
Gunz, Philipp, \& Mitteroecker, Philipp. 2013.
\newblock Semilandmarks: a method for quantifying curves and surfaces.
\newblock {\em Hystrix, the Italian journal of mammalogy}, {\bf 24}(1),
  103--109.

\bibitem[\protect\citename{Guti{\'e}rrez {\em et~al.},
  }2019]{gutierrez2019bayesian}
Guti{\'e}rrez, Luis, Guti{\'e}rrez-Pe{\~n}a, Eduardo, Mena, Rams{\'e}s~H, {\em
  et~al.} 2019.
\newblock A Bayesian Approach to Statistical Shape Analysis via the Projected
  Normal Distribution.
\newblock {\em Bayesian Analysis}, {\bf 14}(2), 427--447.

\bibitem[\protect\citename{Helmert, }1876]{helmert1876genauigkeit}
Helmert, FR. 1876.
\newblock Die Genauigkeit der Formel von Peters zur Berechnung des
  wahrscheinlichen Beobachtungsfehlers director Beobachtungen gleicher
  Genauigkeit.
\newblock {\em Astronomische Nachrichten}, {\bf 88}, 113.

\bibitem[\protect\citename{Kendall, }1977]{kendall1977diffusion}
Kendall, David~G. 1977.
\newblock The diffusion of shape.
\newblock {\em Advances in applied probability}, {\bf 9}(3), 428--430.

\bibitem[\protect\citename{Klingenberg {\em et~al.},
  }2002]{klingenberg2002shape}
Klingenberg, Christian~Peter, Barluenga, Marta, \& Meyer, Axel. 2002.
\newblock Shape analysis of symmetric structures: quantifying variation among
  individuals and asymmetry.
\newblock {\em Evolution}, {\bf 56}(10), 1909--1920.

\bibitem[\protect\citename{Lancaster, }1965]{lancaster1965helmert}
Lancaster, HO. 1965.
\newblock The helmert matrices.
\newblock {\em The American Mathematical Monthly}, {\bf 72}(1), 4--12.

\bibitem[\protect\citename{Lehmann \& Casella, }2006]{lehmann2006theory}
Lehmann, Erich~L, \& Casella, George. 2006.
\newblock {\em Theory of point estimation}.
\newblock Springer Science \& Business Media.

\bibitem[\protect\citename{Mahalanobis, }1925]{mahalanobis1925analysis}
Mahalanobis, Prasanta~C. 1925.
\newblock Analysis of race-mixture in Bengal.

\bibitem[\protect\citename{Mahalanobis, }1936]{mahalanobis1936generalized}
Mahalanobis, Prasanta~Chandra. 1936.
\newblock On the generalized distance in statistics.
\newblock National Institute of Science of India.

\bibitem[\protect\citename{Mardia {\em et~al.}, }2000]{mardia2000statistical}
Mardia, Kanti~V, Bookstein, Fred~L, \& Moreton, Ian~J. 2000.
\newblock Statistical assessment of bilateral symmetry of shapes.
\newblock {\em Biometrika},  285--300.

\bibitem[\protect\citename{Masel, }2012]{masel2012rethinking}
Masel, Joanna. 2012.
\newblock Rethinking Hardy--Weinberg and genetic drift in undergraduate
  biology.
\newblock {\em BioEssays}, {\bf 34}(8), 701--710.

\bibitem[\protect\citename{Micheas \& Peng, }2010]{micheas2010bayesian}
Micheas, Athanasios~C, \& Peng, Yuqiang. 2010.
\newblock Bayesian Procrustes analysis with applications to hydrology.
\newblock {\em Journal of Applied Statistics}, {\bf 37}(1), 41--55.

\bibitem[\protect\citename{Micheas {\em et~al.}, }2006]{micheas2006complex}
Micheas, Athanasios~C, Dey, Dipak~K, \& Mardia, Kanti~V. 2006.
\newblock Complex elliptical distributions with application to shape analysis.
\newblock {\em Journal of statistical planning and inference}, {\bf 136}(9),
  2961--2982.

\bibitem[\protect\citename{O'Higgins \& Dryden, }1993]{o1993sexual}
O'Higgins, Paul, \& Dryden, Ian~L. 1993.
\newblock Sexual dimorphism in hominoids: further studies of craniofacial shape
  differences in Pan, Gorilla and Pongo.
\newblock {\em Journal of Human Evolution}, {\bf 24}(3), 183--205.

\bibitem[\protect\citename{Perez {\em et~al.}, }2006]{perez2006differences}
Perez, S~Ivan, Bernal, Valeria, \& Gonzalez, Paula~N. 2006.
\newblock Differences between sliding semi-landmark methods in geometric
  morphometrics, with an application to human craniofacial and dental
  variation.
\newblock {\em Journal of anatomy}, {\bf 208}(6), 769--784.

\bibitem[\protect\citename{Rohlf \& Slice, }1990]{rohlf1990extensions}
Rohlf, F~James, \& Slice, Dennis. 1990.
\newblock Extensions of the Procrustes method for the optimal superimposition
  of landmarks.
\newblock {\em Systematic Biology}, {\bf 39}(1), 40--59.

\bibitem[\protect\citename{Stange {\em et~al.}, }2018]{stange2018study}
Stange, Madlen, Aguirre-Fern{\'a}ndez, Gabriel, Salzburger, Walter, \&
  S{\'a}nchez-Villagra, Marcelo~R. 2018.
\newblock Study of morphological variation of northern Neotropical Ariidae
  reveals conservatism despite macrohabitat transitions.
\newblock {\em BMC evolutionary biology}, {\bf 18}(1), 38.

\bibitem[\protect\citename{Stern, }1943]{stern1943hardy}
Stern, Curt. 1943.
\newblock The hardy-weinberg law.
\newblock {\em Science}, {\bf 97}(2510), 137--138.

\bibitem[\protect\citename{Theobald \& Wuttke, }2006]{theobald2006empirical}
Theobald, Douglas~L, \& Wuttke, Deborah~S. 2006.
\newblock Empirical Bayes hierarchical models for regularizing maximum
  likelihood estimation in the matrix Gaussian Procrustes problem.
\newblock {\em Proceedings of the National Academy of Sciences}, {\bf 103}(49),
  18521--18527.

\end{thebibliography}


\end{document}